# Active wetting/de-wetting of focal adhesions on viscoelastic substrates

Ivana Pajic-Lijakovic [1,5], Milan Milivojevic [1], Boris Martinac [2,3] Massimo Vassalli [4] and Peter V. E. McClintock[5]

[1] University of Belgrade, Faculty of Technology and Metallurgy, Department of Chemical Engineering, Belgrade, Serbia

[2] Mechanosensory Biophysics Laboratory, Victor Chang Cardiac Research Institute, Sydney, Australia

[3] St Vincent's Clinical School, Faculty of Medicine, University of New South Wales, Sydney, New South Wales, Australia

[4] University of Glasgow, Centre for the Cellular Microenvironment, Glasgow, UK

[5] Department of Physics, Lancaster University, Lancaster LA1 4YB, UK

Correspondence to: Ivana Pajic-Lijakovic, iva@tmf.bg.ac.rs and,

Peter V. E. McClintock, p.v.e.mcclintock@lancaster.ac.uk

**Abstract**

Cell adhesion to viscoelastic substrates is mediated by focal adhesions (FAs), which dynamically couple actomyosin contractility to the extracellular matrix. Although substrate stress relaxation is known to regulate adhesion stability and cell migration, a predictive physical framework linking viscoelasticity to force transmission and adhesion dynamics remains lacking.

Here we review briefly what is known about the active wetting and de-wetting of FAs on viscoelastic substrates and synthesize existing experimental and theoretical work into a two-timescale physical framework to describe the phenomena reported. At short timescales, oscillatory actomyosin-driven displacements are transmitted through molecular clutches, leading to frequency-dependent energy transfer to the substrate. We show that this transfer is maximized at an optimal frequency set by a balance between elastic energy storage and viscous dissipation, establishing a resonance-like mechanism that selects both the effective FA stiffness and traction force amplitude. At longer timescales, this mechanically optimal state couples to adhesion remodelling through an effective surface tension, enabling FA growth and disassembly to be interpreted as active wetting and de-wetting processes. The model predicts that adhesion stability and steady-state size are controlled by substrate stiffness and viscoelastic timescales, as well as mechanosensitive feedback mediated by Piezo1-dependent calcium signalling.

**Key words**: molecular clutches substrate viscoelasticity, integrin-ligand adhesion, cohesion of focal adhesions, actomyosin oscillations

## 1.Introduction

Cell adhesion to the extracellular matrix (ECM) is a fundamental process underlying migration, morphogenesis, and tissue homeostasis [1-4]. The process is governed by focal adhesions (FAs), dynamic mechano-sensitive complexes that transmit forces generated by the actomyosin cytoskeleton to the substrate [5]. Through this mechanical coupling, cells continuously probe their microenvironment and regulate adhesion growth, signalling, migration and lineage specification in the case of stem cells [4,6, 7]. While classical studies have established the role of substrate stiffness in these processes, it is now evident that most biological matrices are viscoelastic, exhibiting time-dependent stress relaxation that critically affects cell behavior [2, 8,9].

Recent experiments have demonstrated that substrate viscoelasticity regulates FA size, lifetime, and force transmission, as well as downstream processes such as cell spreading and mechano-transduction [2-4]. However, despite extensive experimental evidence, a predictive physical framework connecting substrate viscoelasticity to the dynamic assembly and stability of FAs is still lacking. How cells integrate time-dependent mechanical cues with their intrinsic active dynamics remains an open question.

A key feature of cell–matrix interaction is the presence of multiple intrinsic timescales. At short-times (on the order of tens of seconds), actomyosin contractility generates oscillatory forces that are transmitted to the substrate through molecular clutches, which consist of: integrins, adaptor proteins (talin and vinculin) and the actin cytoskeleton. The molecular clutch model was initially proposed by Chan and Odde (2008) to provide a mechanistic understanding of how cells detect and react to the mechanical characteristics of the ECM [10].

At longer-times (minutes), FAs remodel through changes in clutch binding, adhesion area, and mechanosensitive signaling pathways, including the activation of ion channels such as Piezo1 and the associated calcium influx (2,11,12]. Recently, a connection has been established between Piezo1 and integrin-mediated FA signalling, suggesting that the activity of the Piezo1 channels serves as a crucial mediator of integrin signalling, thus influencing cell-ECM interactions [11,13]. The interplay between these fast mechanical processes and slow biochemical adaptation is central to adhesion dynamics but remains poorly understood.

The alteration in the structure of FAs throughout maturation and disassembly of FAs is influenced by the adhesive forces between integrins and ligands, which are stronger than cohesive forces between the FAs that contract into isolated foci [14]. This phenomenon can be regarded as corresponding to active processes of wetting and de-wetting [14]. When adhesion dominates, FA components extend along the substrate, resembling a wetting process; when cohesion and tension dominate, FAs retract and fragment, analogous to de-wetting. These processes play a crucial role in the dynamics occurring at different interfaces of soft-matter systems [15]. Although the active processes of wetting and de-wetting have been examined at a supracellular level in relation to epithelial spreading on substrate matrices [16-18] this phenomenon is not well explored in relation to the molecular dynamics associated with the growth and stability of FAs.

Taken together, these observations point toward the need for a unified physical framework that integrates: (i) viscoelastic substrate response, (ii) active force generation by the cytoskeleton, and (iii) adhesion growth and stability governed by interplay between adhesion and cohesion properties. Recent theoretical efforts have begun to address these aspects by incorporating viscoelastic substrate models, such as the standard linear solid (Zener) model, into descriptions of clutch-mediated force transmission. These approaches suggest that the interplay between elastic energy storage and viscous dissipation can give rise to characteristic timescales or frequencies that regulate force transmission efficiency. Such behavior has been interpreted in terms of resonance-like mechanisms, where optimal mechanical coupling occurs when cellular activity matches substrate relaxation dynamics [4,19].

Building on these ideas, an emerging perspective is to couple short-timescale mechanical optimization with long-timescale structural evolution of adhesions. Within this view, FA growth and shrinkage can be interpreted as active wetting and de-wetting processes, governed by the balance between cytoskeletal forcing, substrate dissipation, and effective adhesion cohesion. This framework provides a potential route to link timescale matching with adhesion stability and mechanosensitive responses.

In this review, we synthesize current experimental and theoretical advances on cell adhesion to viscoelastic substrates, with a particular focus on the role of timescale-dependent mechanics. We discuss how oscillatory force generation, substrate relaxation, and adhesion energetics can be integrated into a coherent physical picture, and highlight open questions regarding the coupling between mechanical and biochemical processes in FA dynamics.

## 2. Structural organization of focal adhesion

A single FA is composed of thousands of molecular clutches that link the actomyosin cytoskeleton to the substrate matrix [10]. However, most of them are inactive. The density of active molecular clutches is influenced by the stiffness of the substrate matrix, typically ranging from 50 to 100 $\mu m^{-2}$ [20]. Each FA contains hundreds to thousands of talin proteins per $\mu m^2$ [21]. A schematic presentation of a matured FA is shown in **Figure 1**:

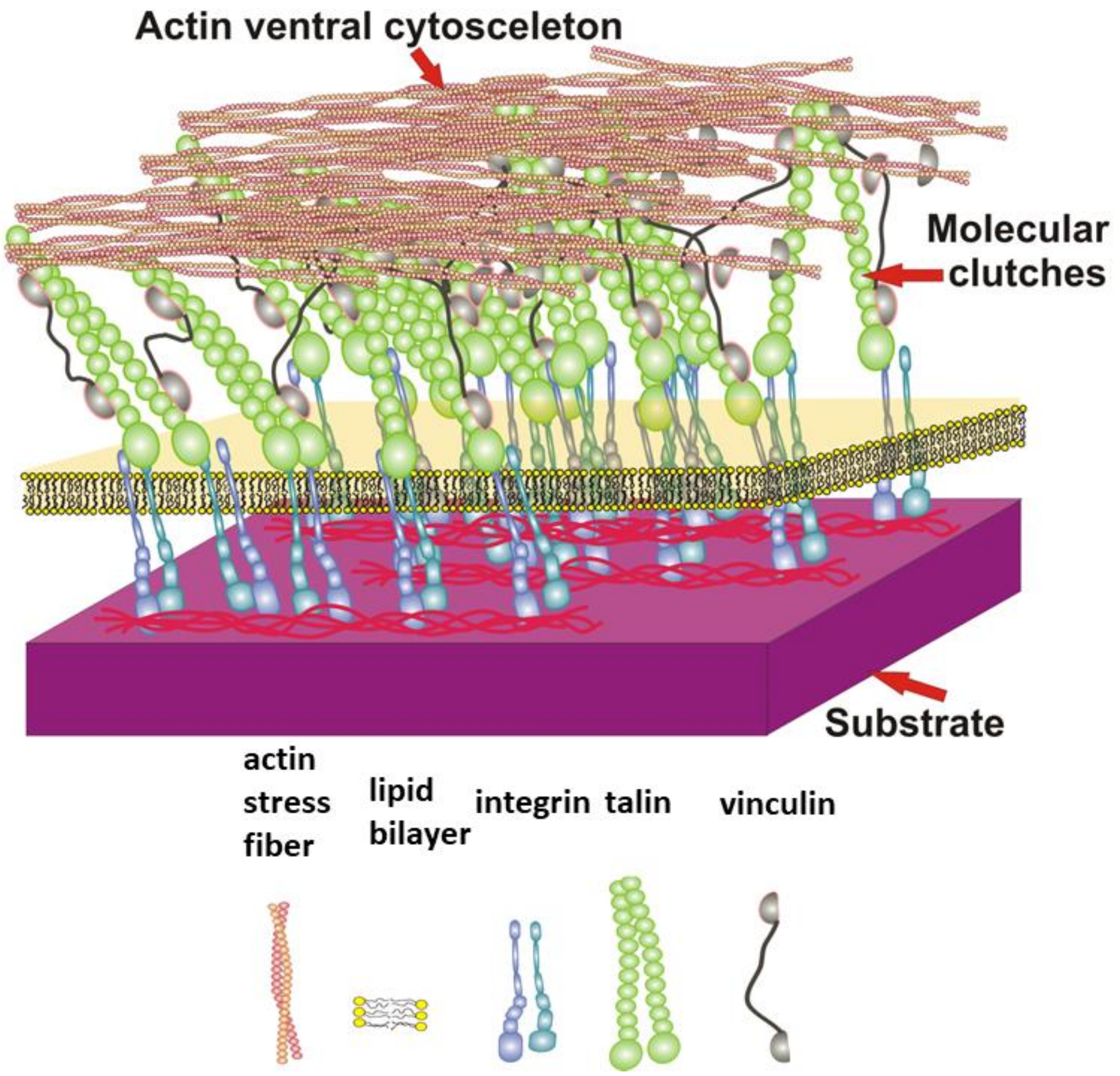

**Figure 1**. A schematic representation of a mature FA on a viscoelastic substrate matrix, incorporating solely the primary factors that contribute to the adhesion and cohesion of the FA. Remodeling of FA is altered by coupling between three layers: (i) actin cytoskeleton, (ii) ensemble of molecular clutches, and (iii) viscoelastic substrate matrix.

Nevertheless, only a minor proportion of these proteins participate in active clutches. Talin exhibits one of the slowest turnover rates among FA proteins, with recovery of a fluorescence half-time of roughly 49.4 seconds [22]. Although talin possesses 11 potential vinculin-binding sites (VBSs), only a limited subset of them is occupied at any given time, leading to an average of approximately 2 vinculin molecules per talin in mature FAs [23]. Vinculin functions as a "molecular lock," inhibiting the refolding of talin [24]. As long as vinculin remains attached to the vinculin binding sites (VBS) the talin rod domain is physically obstructed from refolding into its original helical bundle. If mechanical tension is alleviated (for instance, by inhibiting myosin), the complex retains stability for about 30 minutes before vinculin detaches from the adhesion site [25]. Once the vinculin-talin complex ultimately dissociates, the talin rod fragment can revert to its native configuration in less than one minute under very low forces. Each individual 'clutch' is composed of a single integrin heterodimer that is bound to an extracellular ligand and internally anchored to a single talin molecule [26]. Integrins are arranged in nano-clusters that are connected to substrate ligands, talin molecules, and other FA proteins. Nevertheless, only a minor proportion of integrin molecules actively withstand force. The spatial organization of extracellular ligands at the nanoscale represents an additional regulatory layer in FA mechanosensing. Recent studies have demonstrated that ligand distribution and spacing modulate integrin clustering, molecular clutch engagement, and downstream actin organization, thereby regulating the efficiency of force transmission toward the nucleus [27]. Thus, the adhesive interface is defined not only by ligand availability but also by its spatial organization, which determines the effective density and connectivity of active molecular clutches. Within the present framework, ligand spacing can regulate the local clutch density and effective focal adhesion stiffness, thereby influencing FA wetting, maturation, and mechanochemical adaptation. Consequently, the effective adhesion energy governing FA spreading depends not only on ligand chemistry but also on ligand nanoscale spacing and the resulting organization of mechanically engaged clutches. Additional crucial proteins within the clutch include: (i) kindlin, which is vital for the activation of integrins, (ii) actinin, serving as a secondary link between talin and actin, (iii) FAK (focal adhesion kinase), functioning as a signalling regulator, (iv) paxillin, recognized as a primary "scaffold" protein, and (v) zyxin, which is specifically recruited under conditions of high tension and serves as a reinforcement protein that aids in the repair or fortification of the actin-clutch interface when subjected to significant mechanical load [28-30].

Lateral connection between clutches is essential for the establishment of cohesion within the FA necessary for its stabilization. The most direct 'bridge' between two clutches is the talin-talin C-terminal dimer ($K_D \sim 1.8 - 3\ \mu\mathrm{M}$) [31,32]. Force-induced unfolding of talin molecules ensures vinculin binding to the talin rod [19-33]. Vinculin molecules possess a self-association tendency, leading to the establishment of connections between pairs of clutches. These pairs can also establish strong connections to the same actin stress fiber. Furthermore, integrins often partition into specific lipid domains (rafts), creating lateral cohesion through the physical properties of the lipid bilayer itself. The concentration of proteins like kindlin and paxillin is so high within the FA that they create a 'sticky' environment, which contributes to the internal cohesion. This interconnection is what prevents a single 'clutch' from failing individually; if one integrin-talin bond breaks, the vinculin meshwork redistributes the load to neighbouring clutches.

In a matured FA, the clutch binds more strongly to the substrate (integrin-ligand) than to the actin stress fiber. These active integrin-ligand bonds can withstand forces of $40 - 100\ \mathrm{pN}$ [34]. In contrast, the internal cohesion energy of the FA is a bulk or network property; while individual protein-protein links may be weaker than the integrin-ECM anchor, the collective connectivity of the plaque makes it highly resistant to mechanical extension of the FA. This hierarchy ensures that the FA remains anchored to the substrate even as the internal protein network remodels under tension.

Active wetting and de-wetting arise naturally as the physical mechanisms responsible for FA growth and disassembly. The FA spreads over a wider area when integrin-ligand adhesion energy per unit surface area is higher than the internal cohesion energy of the FA, represented in the form of the spreading factor. While adhesion resists sliding on the substrate, cohesion resists the tearing apart of the FA itself. As the FA extends, it stimulates the activation of new clutches, which subsequently undergo stretching and tilting as shown in **Figure 2**:

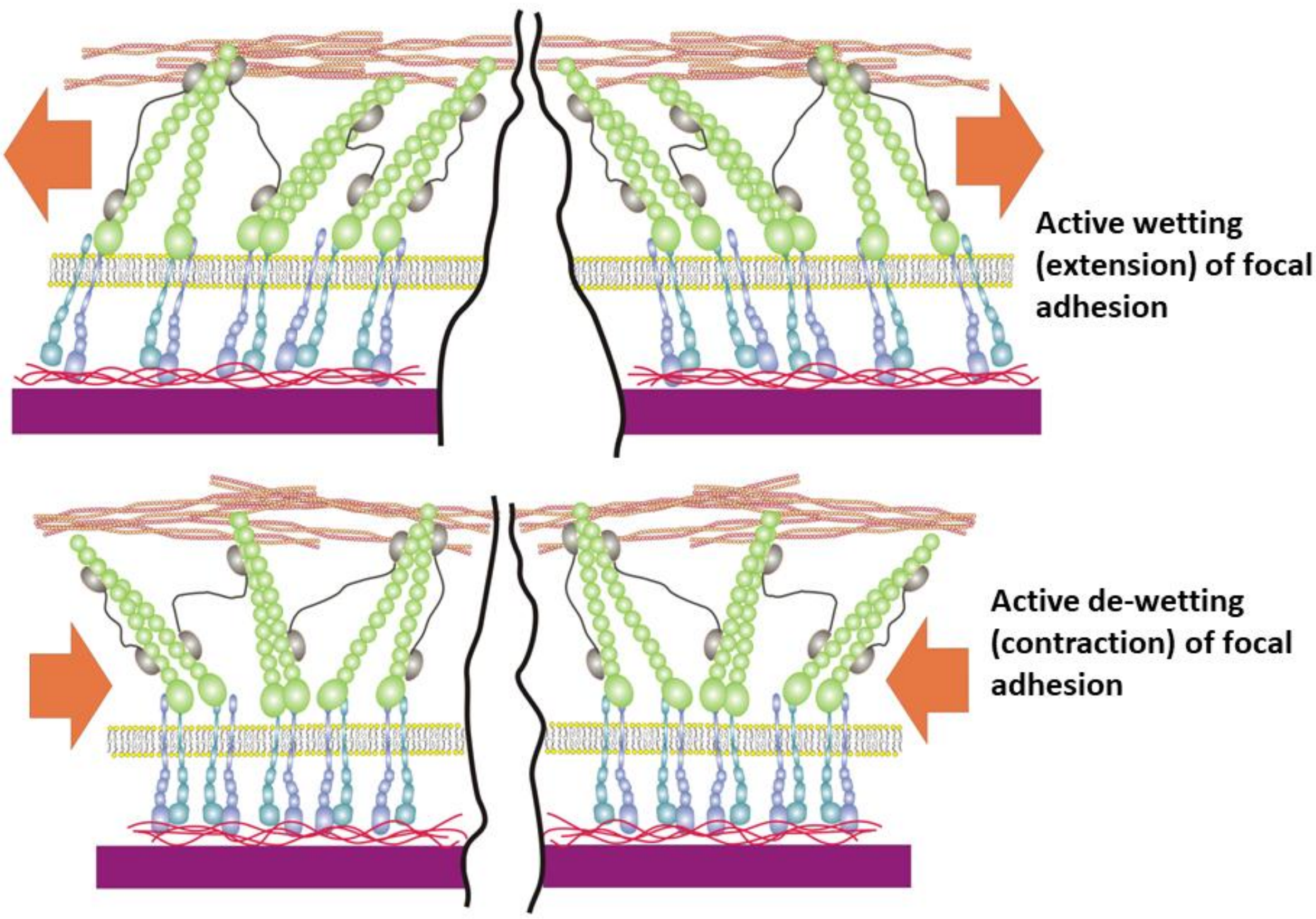

**Figure 2**. The stretching and tilting of molecular clutches during FA wetting results in the unfolding of talin and the recruitment of vinculin, which serves to strengthen the internal protein network, thereby enhancing the stability and cohesion of FA. During the process of wetting, molecular clutches experience both stretching and tilting while maintaining their alignment. An increase in cohesion energy relative to adhesion energy, leading to a spreading factor that is less than zero, causes the de-wetting of the FA. The disruption of clutch alignment and the conformational alterations that occur during the de-wetting process contribute to disassembly of the FA. Orange arrows represent the direction of FA deformation, i.e., extension or contraction.

The tilting of clutches stimulates the maturation of the FAs. This mechanical strain recruits vinculin to reinforce the internal protein network, increasing the internal cohesion energy. However, if the tension exceeds the peak of the integrin catch-bond, or leads to bond rupture, the adhesion energy drops relative to the cohesive tension of the actomyosin network. This initiates the de-wetting transition, leading to contraction of the FA and, as a result, disassembly occurs as the internal network reallocates the load.

### 2.1 Role of Piezo1 channels in maintaining the stability of FAs

The distribution and activity of Piezo1 channels play a key role in regulating focal adhesion (FA) dynamics. Piezo1 is relatively uniformly distributed in mesenchymal-like cancer cells. In epithelial cells, its distribution is more heterogeneous [11-12]. In these cells, Piezo1 accumulates and clusters around FAs, where it more directly influences adhesion dynamics [11]. This clustering is enhanced by actomyosin contractility [11]. Two main mechanisms control Piezo1 distribution along the cell membrane. First, Piezo1 undergoes lateral diffusion driven by concentration gradients. In the absence of other driving forces, this process leads to a uniform distribution [35]. Second, membrane curvature develops at FAs. This effect is stronger in epithelial cells due to the more homogeneous organization of ventral stress fibers compared to cancer cells [12]. Curvature changes local membrane area and generates membrane surface tension gradients. These gradients induce the Marangoni-like flows of Piezo1 channels. As a result, Piezo1 channels are redistributed and accumulated at FAs [11-12]. The activity of Piezo1 channels depends on the spatial regularity of membrane curvature [36]. Rough inward curvature around focal adhesions allows a fraction of clustered Piezo1 channels to retain their preferred outward-facing footprints, whereas smooth curvature promotes an outward-to-inward footprint transition, enabling the entire Piezo1 cluster to undergo cooperative pit-like invagination within the curved membrane region [37].

In the context of active wetting and de-wetting, mechanosensitive Piezo1 channels act as the primary "mechanostat'' or regulator of the spreading factor. Piezo1 plays a dual role: (i) promoting clutch engagement and FA extension during the wetting phase, and (ii) acting as a mechanochemical trigger for the transition from wetting to de-wetting by generating localized $Ca^{2+}$ signals that bias FA turnover pathways, thereby promoting the selective disassembly and recycling of over-stressed or mechanically inefficient clutches [35]. When the FA becomes too large or the tension too high, the collective activation of Piezo1 channels reaches a threshold where calcium entry becomes excessive [35,38]. Elevated local calcium biases adhesion dynamics toward turnover by activating calcium-sensitive pathways, including calpain-mediated cleavage of focal adhesion components such as talin and vinculin, progressively reducing clutch stability and lifetime [11,39]. In the presence of sustained actomyosin tension, the alterations in turnover kinetics promote the retraction and fragmentation of the adhesion, resembling an active de-wetting response [40].

Because FAs are transient, non-equilibrium assemblies, the balance between adhesion-driven spreading and contractility-driven retraction is inherently time-dependent. In stress-relaxing, natural ECMs, the mechanical resistance encountered by tilting clutches decays over time, reducing the rate at which tension accumulates on integrin–ligand bonds, biasing adhesion dynamics toward longer lifetimes by sustaining effective force transmission and favoring a spreading-dominated regime for extended periods (spreading factor $S_{FA} > 0$). Over time, cumulative mechanochemical signalling associated with force transmission, including Piezo1-mediated $Ca^{2+}$ influx, gradually shifts focal adhesion turnover towards disassembly. Piezo1-mediated $Ca^{2+}$ entry activates calcium-dependent signalling pathways that weaken the molecular cohesion of adhesion complexes, while also promoting actomyosin contractility through $Ca^{2+}$-dependent myosin activation. The increased contractile forces then act on these mechanically weakened adhesions, accelerating their rupture and turnover. In the presence of sustained actomyosin tension, this shift biases the system toward retraction and fragmentation, corresponding to an active de-wetting-like response ($S_{FA} < 0$) that marks the conclusion of the FA life cycle. The viscoelastic properties of the substrate matrix affect the stability and remodeling of focal adhesions (FA), in addition to influencing the rate at which FA becomes wetted. In this statement, it is essential to examine the coupling between the FA and the viscoelastic substrate by identifying the primary physical parameters that govern this process.

**3. Coupling of focal adhesions to viscoelastic substrate matrices: the short-time dynamics**

The dynamics of an FA depends on a series-coupling of three elements: (i) the actin cytoskeleton of the cell, (ii) the ensemble of active molecular clutches, and (iii) the substrate matrix [10]. A molecular clutch is a dynamic biochemical linkage that transmits contractile force from the actin cytoskeleton to the extracellular matrix to regulate cell signalling and cell movement. The FA-substrate coupling is considered on two timescales. The short-time scale, measured in seconds, is associated with individual actomyosin oscillations, whereas the long-time scale, which is measured in minutes, pertains to the maturation of FAs. The traction force at an FA is expressed as [4,10]:

$$F(r,t,\tau) = \langle f_c(r,t)\rangle_{t_s} n(r,\tau)\Delta A(r,\tau) \quad (1)$$

where $t$ is the short timescale, $\tau$ is the long timescale, $r$ is the local coordinate, $\langle f_c(r,t)\rangle_{t_s}$ is the course grained force of single clutch equal to: $\langle f_c(r,t)\rangle_{t_s} = \frac{1}{\Delta t_s}\int_t^{t+\Delta t_s} f_c(r,t)dt$, $\Delta t_s$ is the short time increment that satisfies the condition that $\Delta t_s \ll T$, $T$ is the period of actomyosin oscillations, $f_c(r,t) = \frac{\Delta A}{N}\sum_i^N f_{ci}\delta(r-r_i)$ is the space coarse grained force per single clutch, $f_{ci}$ is the force of the i-th clutch, $N$ is the number of active clutches, $\Delta A(r,\tau)$ is the increment of FA surface area, and $n(r,\tau) = \sum_i^N \delta(r-r_i)$ is the surface density of active molecular clutches, which depends on the substrate stiffness and can be estimated as lying in the range of $10-100\ \frac{clutch}{\mu m^2}$ such that higher values of $n$ appear on stiffer substrates [40]. The surface area of a matured FA is in the range $0.5-5\ \mu m^2$ depending on the stiffness of substrate matrix [41]. All elements experience the same coarse grained traction force in space and time [4]:

$$F(r,t,\tau) = F_a(r,t,\tau) = F_{FA}(r,t,\tau) = F_m(r,t,\tau) \quad (2)$$

where $F_a(r,t,\tau)$ is the force exerted by actin stress fibers connected to the talin cluster, $F_{FA}(r,t,\tau)$ is the force exerted by the ensemble of molecular clutches from the cytoskeleton, and $F_m(r,t,\tau)$ is the force exerted by the substrate matrix from the cytoskeleton through the molecular clutches. Cells cultured on stiffer gels generate significantly greater traction forces. Conversely, cells on softer gels induce larger physical displacements (strains) within the material, as the substrate is more readily deformed [42]. The tangential displacements of these elements in the direction opposite to that of cell migration [43] satisfy the condition that [4]:

$$u_a(r,t,\tau) = u_m(r,t,\tau) + u_{FA}(r,t,\tau) \quad (3)$$

where $u_m(r,t,\tau)$ is the substrate displacement field caused by actomyosin contractions, $u_{FA}(r,t,\tau)$ is the displacement of the ensemble of clutches, and $u_a(r,t,\tau)$ is the displacement of actin stress fibers connected to talin molecules during actomyosin contractions. The displacement $u_a(r,t,\tau)$ is oscillatory and can be expressed as:

$$u_a(r,t,\tau) = u_0(r,\tau)\sin(\omega(r)t) \quad (4)$$

where $u_0(r,\tau)$ is the amplitude of actomyosin oscillations, which varies along FA depending on cytoskeletal, adhesive, and substrate properties, and $\omega(r)$ is the angular velocity equal to $\omega(r) = \frac{2\pi}{T(r)}$, and $T(r) \sim 30 - 90\ \mathrm{s}$ is the period of the oscillations [44]. The amplitude $u_0(r,\tau)$ depends on the intracellular concentration of calcium $[C_a^{2+}]$. Calcium binds to calmodulin, which activates myosin light chain kinase (MLCK) [45,46]. This increases the number of active myosin motors pulling on the actin filaments leading to an increase in the amplitude $u_0(r,\tau)$. The amplitude of actomyosin oscillations is $u_0 = 50 - 350\ \mathrm{nm}$, while expression of the talin-binding head of vinculin increases stretching to about 400 nm [47]. The number of vinculin is higher than the number of talin per single FA [48). While talin acts as the primary scaffold linking integrins to actin, it contains up to 11 vinculin-binding sites (VBS) within its rod domain. This structural feature allows a single talin molecule to recruit and bind multiple vinculin molecules simultaneously once tension is applied [49].

The FA behaves elastically at a short timescale and viscoelastic at a long timescale. The short-time displacement of the ensemble of active clutches is equal to [4,10]:

$$u_{FA}(r,t,\tau) = \frac{1}{k_{FA}(r,\tau)} F(r,t,\tau) \quad (5)$$

where $k_{FA}$ is the FA spring constant that resists pulling by the actomyosin and is equal to $k_{FA}(r,\tau) = n(r,\tau)\Delta A(r,\tau)k_{cs}$ (where $k_{cs}$ is the average spring constant of a single clutch). The spring constant $k_{cs}$ can be expressed by including all resistance effects as: $\frac{1}{k_{cs}} = \frac{1}{k_{actin-talin}} + \frac{1}{k_{talin}} + \frac{1}{k_{integrin}} + \frac{1}{k_{linkers}}$ (where $k_{actin-talin}$, $k_{talin}$, $k_{integrin}$, $k_{linkers}$ are the spring constants of the various parts of the molecular clutch). Integrins are generally stiffer than talin although they still contribute to the overall flexibility. Talin molecules become softer during unfolding. The spring constant of folded talin $k_{talin}^f$ is in the range $k_{talin}^f = 0.5 - 1\ \frac{\mathrm{pN}}{\mathrm{nm}}$, while the spring constant of unfolded talin $k_{talin}^{uf}$ is in the range $k_{talin}^{uf} = 0.01 - 0.1\ \frac{\mathrm{pN}}{\mathrm{nm}}$ [50]. When vinculin attaches to an unfolded talin molecule and stabilises it, talin retains its unfolded state for more than $100\ \mathrm{s}$, even when the mechanical tension is removed [51]. The spring constant of FA $k_{FA}(r,\tau)$ can be estimated to lie in the range $1 - 100\ \frac{\mathrm{pN}}{\mathrm{nm}}$ depending on the substrate stiffness [51].

Oscillation of actomyosin causes an oscillation of the traction force [52]:

$$F(r,t,\tau) = F_0(r,\tau)\sin\big(\omega(r)t + \varphi(r,\tau)\big) \quad (6)$$

where $F_0(r,\tau)$ is the local amplitude of oscillation and $\varphi(r,\tau)$ is the phase lag in the respect of the amplitude of the actomyosin displacement $u_0(r,\tau)$. The magnitude of the amplitude $F_0$ must reach the range over which talin rod domains begin to unfold (~5–25 pN), thereby exposing vinculin-binding sites and enabling progressive FA reinforcement [53].

Through the analysis of experimental data pertaining to various substrate matrices found in the literature [2-4], we suggest the Zener model (suitable for viscoelastic solids) as a fitting general model for elucidating the viscoelastic behaviour of substrate matrices. This model encompasses the relaxation of force (stress) and displacement (strain) of substrates induced by cellular traction. As a result, the Zener constitutive model representing the relationship between force and displacement of substrates was articulated as [4]:

$$F(r,t,\tau) + \tau_m(r)\frac{dF(r,t,\tau)}{dt} = k_m(r)u_m(r,t,\tau) + k_m(r)\tau_u(r)\frac{du_m(r,t,\tau)}{dt} \qquad (7)$$

where $\tau_m(r)$ is the force (substrate stress) relaxation time, $k_m(r)$ is the equilibrium spring constant of the substrate, which varies along the FA surface area, and $\tau_u(r)$ is the relaxation time of the displacement $u_m$ (i.e., the retardation time), which satisfies the condition that $\tau_u > \tau_m$ [54]. This linear formulation should be interpreted as an effective approximation of experimentally characterized cell–matrix systems obtained on biomimetic substrates, including PEG-based hydrogels, polyacrylamide (PAAm) substrates, and alginate–reconstituted basement membrane (rBM) systems, rather than a universal description of extracellular matrix mechanics. More complex matrices, such as fibrous collagen networks, exhibit hierarchical relaxation spectra and nonlinear responses that may require extensions involving multiple relaxation modes and deformation-dependent material parameters. The relationship between the relaxation and retardation times for homogeneous and isotropic polymer gels is in the range of $\frac{\tau_m}{\tau_u} = 0.3 - 0.8$ and depends on the mobility of the polymer chains, while the mobility of the chains depends on the substrate stiffness expressed in terms of the spring constant $k_m$ or Young's modulus $E_m$. The relaxation time $\tau_m$ relies on the strength of the inter-chain bonds, while the retardation time $\tau_u$ depends on the mobility of the chains [55]. Higher chain mobility, characteristic of softer substrates, results in a decrease in the ratio between the two relaxation times, i.e., $\frac{\tau_m}{\tau_u} \sim k_m$. The spring constant of the substrate can be related to the Young's modulus $E_m$ of a homogeneous and isotropic substrate, i.e., $k_m = \frac{2R_{FA}E_m}{1-\nu_m^2}$ (where $R_{FA}$ is the radius of FA equal to $R_{FA} = 0.5 - 2$ μm and $\nu_m$ is the Poisson's ratio of the substrate measuring the deformation of the substrate in direction perpendicular to the direction of the applied force) [56].

The introduction of: (i) $u_m(r,t,\tau) = u_0(r,\tau)\sin(\omega(r)t) - \frac{F(r,t,\tau)}{k_{FA}(r,\tau)}$ and (ii) $\frac{du_m(r,t,\tau)}{dt} = u_0(r,\tau)\omega\cos(\omega(r)t) - \frac{1}{k_{FA}(r,\tau)}\frac{dF(r,t,\tau)}{dt}$ into eq. 7, establishes the $F$-$u_a$ relationship:

$$F\left(1+\frac{k_m}{k_{FA}}\right) + \frac{dF}{dt}\tau_m\left(1+\frac{k_m}{k_{FA}}\right) = k_m u_0\sin(\omega t) + k_m\tau_u u_0\omega\cos(\omega t) \qquad (8)$$

The force amplitude $F_0$ and phase lag $\varphi$ can be expressed by inserting eq. 6 into eq. 8:

$F_0(k_{FA},\omega) = k_{FA}k_m u_0\sqrt{\frac{1+\omega^2\tau_u^2}{(k_m+k_{FA})^2+\omega^2(k_{FA}\tau_m+k_m\tau_u)^2}}$ and

$$\varphi(k_{FA}, \omega) = \arctan\left[\frac{k_{FA}\omega(\tau_u - \tau_m)}{(k_m + k_{FA}) + \omega^2 \tau_u (k_{FA}\tau_m + k_m \tau_u)}\right]. \tag{9}$$

The magnitude of the force amplitude can be considered within two regimes distinguished from eq. 9: (i) a high-angular velocity regime for $\omega\tau_u \gg 1$and (ii) a low-angular velocity regime for $\omega\tau_u \ll 1$. The force amplitude is equal to: (i) $F_0^{high} = \frac{k_{FA}k_m u_0}{k_m + k_{FA}\frac{\tau_m}{\tau_u}}$ for the high-angular velocity regime and (ii) $F_0^{low} = \frac{k_{FA}k_m u_0}{k_m + k_{FA}}$ for the low-angular velocity regime. Consequently, $F_0^{high} > F_0^{low}$. Viscoelastic energy transfer is maximized within a finite frequency window, highlighting a distinction between force transmission and energy dissipation with important biological implications, since full mechanical synchronisation may suppress rather than enhance mechanosensory excitability.

Upon further examination, it is essential to delineate the function performed by the actomyosin during a single oscillation, specifically the cumulative energy contribution from the cytoskeleton. This work can be expressed as:

$$W(r, \tau) = \int_0^T F(r, t, \tau) \frac{du_a(r, t, \tau)}{dt} dt \tag{10}$$

where the work $W(r, \tau) = \pi F_0(r, \tau) u_0(r, \tau) \sin\varphi(r, \tau)$. The angular velocity range, which ensures maximum work $W_{max}$ can be expressed through the condition: $\frac{dW(\omega)}{d\omega} = 0$, as: $\omega_{max}(r, \tau) = \frac{1 + \frac{k_m}{k_{FA}}}{\tau_m + \tau_u \frac{k_m}{k_{FA}}}$. This is one of the key physical parameter for describing FA-substrate coupling. An increase in the relaxation and retardation times under the same stiffness of the substrate, expressed by the spring constant $k_m$, results in a decrease in the maximum angular velocity $\omega_{max}$. The analysis identifies an optimal range of driving frequencies at which viscoelastic energy transfer from actomyosin to the substrate is maximized. Different regions are tuned to different resonance conditions, evolve at different rates and cannot synchronise globally. When the characteristic timescale of actomyosin activity approaches this mechanically selected frequency, viscoelastic coupling enhances the amplitude and phase-dependent transmission of mechanical forces to the membrane–cytoskeletal interface. This increases the probability that the local mechanical stimulus exceeds the Piezo1 activation threshold, thereby promoting channel opening and calcium influx [57]. To distinguish between the externally transmitted traction force amplitude and the local mechanical stimulus experienced by Piezo1, we introduce a frequency-dependent mechanical transfer function, $\Gamma(\omega)$, that accounts for the viscoelastic filtering and amplification properties of the FA–cytoskeleton–membrane system. The effective force acting on Piezo1 $F_{Piezo1}$ can therefore be expressed as: $F_{Piezo1} = \Gamma(\omega) F_0$. Piezo1 activation occurs when the local force exceeds an effective gating threshold. This resonance-like matching enhances the integrated calcium signal over time, which can support focal adhesion maturation and stability, even on softer substrates [2].

In further consideration, it is essential to define the optimal spring constant of the FA, denoted as $k_{FA}^*$, which guarantees that the maximum force amplitude, $F_0^{max}$, meets the criterion: $\frac{dF_0(k_t, \omega_{max})}{dk_{FA}} = 0$. Consequently, the optimal spring constant of the FA is $k_{FA}^*(r) = k_m \sqrt{\frac{\tau_m}{\tau_u}}$. The difference between the spring constants $k_m(r)$ and $k_{FA}^*(r)$ is lower for softer substrates due to the higher mobility of polymer chains within the substrate. The FA adapts its state to the viscoelasticity of a substrate, described by the physical parameters $k_m(r)$, $\tau_m(r)$, and $\tau_u(r)$ by changing the spring constant $k_{FA}(r,\tau)$ towards its optimal value, i.e., $k_{FA}(r,\tau) \to k_{FA}^*$. With $k_m(r)$ fixed by the substrate and $k_{FA}(r,\tau)$ evolving through active wetting and de-wetting, FAs dynamically scan a spatially heterogeneous resonance landscape. Mechanical coupling through the cytoskeleton prevents simultaneous optimization across all sites, leading to a frustrated state characterized by locally resonant yet globally asynchronous traction dynamics.

### 3.1 Active wetting and de-wetting of FAs: the long-time dynamics

The long-time dynamics of an FA includes changes in the surface density of active clutches $n(r,\tau)$ and in the FA's surface area $\Delta A(r,\tau)$. While the density $n(r,\tau)$ represents an internal order parameter, the surface area $A(\tau)$ represents an external characteristic, i.e., the footprint of the FA.

The lifetime of an FA includes both (i) spreading (maturation) and (ii) contraction. These processes should be discussed in the context of active wetting and de-wetting. They are both governed by a dynamic spreading factor $S_{FA} = S_{FA}(k_{FA}, n)$ which can be expressed as: $S_{FA} = e_a - e_{coh}$ (where $e_a$ is the substrate adhesion energy per unit surface and $e_{coh}$ is the cohesion energy of the FA complex) [17]. The adhesion energy is equal to: $e_a \sim \rho_I E_{bI}$ (where $\rho_I$ is the surface density of integrin molecules and $E_{bI}$ is the integrin-ligand bond energy). The adhesion energy $e_a$ can be estimated for the surface density of integrin in the range of $500 - 1500\ \frac{\text{molecules}}{\mu\text{m}^2}$ [58] and the bond energy for higher affinity integrin-ligand complexes in the range of $20 - 25\ k_B T$ [59], such that $e_a \sim 1x10^{-1}\ \frac{\text{mN}}{\text{m}}$. The cohesion energy $e_{coh}$ increases during wetting process.

When $S_{FA} > 0$, the FA experiences wetting (spreading), which results in its maturation. In this scenario, extension of the FA promotes the active recruitment of new clutches to maintain its structural integrity. Stretching and tilting of the clutches recruits vinculin to reinforce the internal protein network, thereby increasing the internal cohesion energy. An increase in cohesion energy relative to adhesion energy results in a decrease in the spreading factor. When the spreading factor is equal to zero ($S_{FA} = 0$), the FA has reached its mature, state. Further extension of the FA leads to rupture of some integrin-ligand bonds and a decrease in adhesion energy. When the spreading factor becomes $S_{FA} < 0$, the FA undergoes de-wetting (contraction). FA contraction causes active crowding, stimulating interactions among clutches and hence their conformational changes, leading to FA disassembly.

The average adhesion energy $\langle e_a \rangle$, which is related to spring constant of FA and the displacement $u_{FA}$, can be expressed as:

$$\langle e_a(r,\tau) \rangle = \frac{1}{2Ak_{FA}} \left[ \frac{1}{T} \int_0^T F(r,t,\tau)^2 dt \right] \quad (11)$$

which is equal to: $\langle e_a(\tau)\rangle = \frac{F_0^2}{4Ak_{FA}}$. Spreading of an FA causes the tilting of peripheral clutches and disruption of the peripheral clutch-substrate bonds, leading to a decrease in adhesion energy relative to cohesion energy. The maximum adhesion energy occurs when the FA spring constant $k_{FA}(r,\tau) \rightarrow k_{FA}^*(r)$ and the force amplitude reaches its maximum $F_{0\,max}$. The average cohesion energy of the FA can be express thermodynamically as:

$$\langle e_{coh}(r,\tau)\rangle = 2\gamma_{FA}(r,\tau) \quad (12)$$

where $\gamma_{FA}(r,\tau)$ is its surface tension in contact with liquid medium) [17]. The cohesion and adhesion energies depend on the mobility of polymer chains within a substrate. Their increased mobility, which is typical of softer substrates, intensifies the interactions among clutches, which destabilize the FA and cause a decrease in the surface tension $\gamma_{FA}(r,\tau)$.

Surface tension is generally regarded as a state variable and can be expressed thermodynamically through the virial equation, which is a function of the surface packing density of the system's constituents [60]. Consequently, the dynamic surface tension of an FA complex $\gamma_{FA}$ depends on the surface packing density of clutches and the level of intracellular calcium, i.e., $\gamma_{FA} = \gamma_{FA}(n,C)$. It can be expressed by modifying the model [60] as:

$$\frac{\gamma_{FA}(r,\tau)}{e_{eff}n} = 1 - B_2 n \quad (13)$$

where $e_{eff}$ is the effective binding energy per single active clutch and $B_2 = B_2(C)$ is the second virial coefficient, which depends on the level of intracellular calcium, and can be expressed as: $B_2(C) = -\frac{1}{2}\int_0^\infty d^2r_{12}\int_0^{2\pi}\left(e^{-\frac{U(r_{12},\theta_{12},C)}{k_BT}} - 1\right)d\theta_{12}$, $r_{12}$ is the distance between two clutches, $\theta_{12}$ is the orientation angle between two neighbouring clutches, and $U(r_{12},\theta_{12},C)$ is the effective mesoscale pair interaction potential between clutches, that includes the contributions of positional interactions $U(r_{12},C)$ and orientational interactions $U(\theta_{12},C)$. While the classical Chan and Odde model [10] effectively captures force-dependent bond kinetics using independent linear springs, our model improves upon this by incorporating a Lennard-Jones potential. This addition accounts for the mesoscale self-organization of the FA at long timescales by representing effective steric exclusion, cooperative clustering, and cytoskeleton-mediated mechanical constraints that regulate clutch remodeling and stabilization under oscillating actomyosin loads. The pair interaction potential $U(r_{12},\theta_{12},C)$ should be interpreted as an effective mesoscale interaction potential that includes direct clutch–clutch interactions as well as constraints imposed by the surrounding actomyosin cytoskeleton and membrane environment. The positional component can be described phenomenologically using a Lennard–Jones potential:

$$U(r_{12},C) = 4\epsilon(C)\left[\left(\frac{d}{r_{12}}\right)^{12} - \left(\frac{d}{r_{12}}\right)^6\right] \quad (14)$$

where $\epsilon(C)$ determines the strength of the interactions and $d$ is the distance at which the potential $U(r_{12},C)$ is zero. We then complement this with an elastic potential for orientational interactions, which accounts for the energy cost of angular deviations as clutches align with the direction of the actomyosin-driven force. The orientational potential can be expressed as:

$$U(\theta_{12}, C) = \frac{1}{2} k_\theta \Delta\theta_{12}{}^2 \quad (15)$$

where $k_\theta(C)$ is the torsional spring constant of a single clutch). During FA extension the potential $U(r_{12}, C)$ is attractive. In this case, clutches tilt but retain alignment such that $\Delta\theta_{12} \rightarrow 0$ and the orientational potential $U(\theta_{12}, C) \rightarrow 0$. In contrast to FA extension, contraction of an FA perturbs the clutches' alignment leading to an increase in the orientational potential $U(\theta_{12}, C)$, while the positional potential $U(r_{12}, C)$ becomes repulsive. Consequently, the virial coefficient $B_2$ satisfies the condition that: $B_2^{exten} < B_2^{compress}$. An increase in the packing density of active clutches and a decrease in the second virial coefficient $B_2$ cause an increase in the surface tension $\gamma_{FA}$, as well as cohesion energy relative to adhesion energy, which results in a decrease in the spreading factor during FA wetting. When the spreading factor turns negative, the FA experiences contraction, which enhances the interactions among clutches, resulting in an increase in $B_2$ and initiating disassembly of the FA during de-wetting. An increase in the intracellular concentration of calcium, caused by simultaneous activation of Piezo1 channels, has the potential to reduce $B_2$, leading in turn to a further decrease in $B_2$ and an increase in the cohesion force between the FA [2]. A decrease in clutch lifetime and stability of the FA causes a decrease in the surface tension $\gamma_{FA}$.

The surface density of active clutches increases during FA spreading, leading to FA reinforcement and consequently an increase in the traction force. However, an increase in the surface density $n$ caused by FA contraction perturbs the clutches' alignment, thereby triggering their disassembly. Consequently, the change of the surface density $n$ can be formulated by modifying the model proposed by Chan and Odde [10] for inhomogeneous distribution of clutches as:

$$\frac{\partial n(r,\tau)}{\partial \tau} + \frac{du_{FA}}{d\tau} n = k_{on}(C)(F_0 - F_c) - k_{off}(C)n \quad (16)$$

where $k_{on}(C)$ is the recruitment rate constant for active clutches, $k_{off}(C)$ is the rate at which active clutches disengage or become inactive, and $F_c$ is the threshold force. The force amplitude should satisfy the condition that $F_0 > F_c$ for FA spreading and an increase in the surface density of clutches. The relaxation time of the surface packing density of clutches $n(r, \tau)$ is $\tau_n \sim \frac{1}{k_{off}}$. Oliva et al. [2] pointed out that Piezo1 activity decreases the clutch dissociation rate $k_{off}$.

The maturation of an FA should be discussed in the context of the spreading factor. Consequently, the change of surface area of an FA caused by FA wetting/de-wetting is expressed as [15]:

$$\frac{d\Delta A(r,\tau)}{d\tau} = k_g(C) S_{FA} \quad (17)$$

where $\Delta A(r, \tau)$ is the FA surface area increment and $k_g(C)$ is the FA growing rate. The quasi-steady state $\Delta A_{ss}$ occurs when $S_{FA} = 0$ and from the adhesion and cohesion energies $\langle e_a(r, \tau) \rangle$ and $\langle e_{coh}(r, \tau) \rangle$ can be expressed as: $\Delta A_{ss}(r) = \frac{F_0^2}{8\gamma_{FA} k_{FA}}$. In this case, $\gamma_{FA}$ can be estimated as:

$\gamma_{FA} \sim \frac{1}{2}\langle e_a(r,\tau)\rangle$, which is equal to $\gamma_{FA} \sim 0.5x10^{-1}\ \frac{\mathrm{mN}}{\mathrm{m}}$. The surface tension of epithelial monolayers is equal to a few $\frac{\mathrm{mN}}{\mathrm{m}}$ [61].

The rate of intracellular calcium concentration caused by opening a fraction of Piezo1 channels within minutes can be expressed as [12]:

$$\frac{dC(r,\tau)}{d\tau} = k_{in} y_p - k_{out} C_{Ca} \tag{18}$$

where $y_p(r,\tau)$ is the fraction of opening Piezo1 channels within a cluster of channels, $k_{in}$ is the specific rate of calcium inflow, and $k_{out}$ is the specific rate of calcium outflow. The long-term change in the fraction of opening Piezo1 channels $y_p(r,\tau)$, as a measure of their activity, can be expressed in the form of a Langevin-type equation as [12]:

$$\frac{dy_p(r,\tau)}{d\tau} = k_{open}(\omega_{max})\left(1 - y_p\right) - k_{close} y_p + \xi_{int} \tag{19}$$

where $k_{open}(\omega_{max})$ is the channels' specific rate of opening equal to: $k_{open}(\omega_{max}) = k_0 e^{-\frac{W(\omega_{max})}{k_B T}}$, $k_0$ is the basal opening rate of a Piezo1 channel in the absence of applied mechanical force from the cytoskeleton, $k_{close}$ is the channels' specific rate of closing, and $\xi_{int}$ is the stochastic force caused by hydrophobic interactions between Piezo1 channels and lipids. Equations (9–13) show that FA wetting and de-wetting are not imposed processes but emerge from the coupled evolution of force transmission, clutch turnover, and Piezo1-mediated calcium signaling, all operating on a slow timescale.

**4. The impact of substrate viscoelasticity on the wetting/de-wetting of focal adhesions**

The maturation of FAs is best achieved when the cell's actual contraction angular velocity $\omega$ approaches $\omega_{max}$, which ensures maximum energy transfer from cytoskeleton to substrate. High stability of FAs is a prerequisite for efficient cell migration. Unstable FAs on soft substrates and very stable FAs on stiff substrates exhibit reduced cell migration. Optimal stability of FAs depends on the stiffness of the substrate matrices and on the level of intracellular calcium and it can be characterized by optimal: (i) cohesion energy and (ii) lifetime range of FAs. The stability of FAs on stiffer substrates is pronounced in comparison with that on softer substrates [2]. However, the stability of FAs on soft substrates can be improved by interplay between an increase in the level of intracellular calcium, and a decrease in the substrate relaxation and retardation times [2]. The intracellular concentration of calcium depends on the activity of Piezo1 channels. The concurrent opening of a cluster of Piezo1 channels relies on the angular velocity $\omega_{max}$, which further depends on the spring constants of the substrate and of the FA, $k_m$ and $k_{FA}$, respectively, as well as on the relaxation and retardation times of the substrate. Consequently, changes in substrate relaxation and retardation times shift the frequency window $\omega_{max}$ at which Piezo1-mediated mechanosignalling is most effective, by altering the viscoelastic phase response of force transmission. Long relaxation and retardation time cause an increase in $\omega_{max}$ and hence reduce the simultaneous opening of a cluster of Piezo1 channels [2]. Consequently, the effectiveness of collective cell migration can be modulated by altering the relaxation and retardation lifetime of substrates [2-4]. The relationships between physical parameters are shown in **Figure 3**:

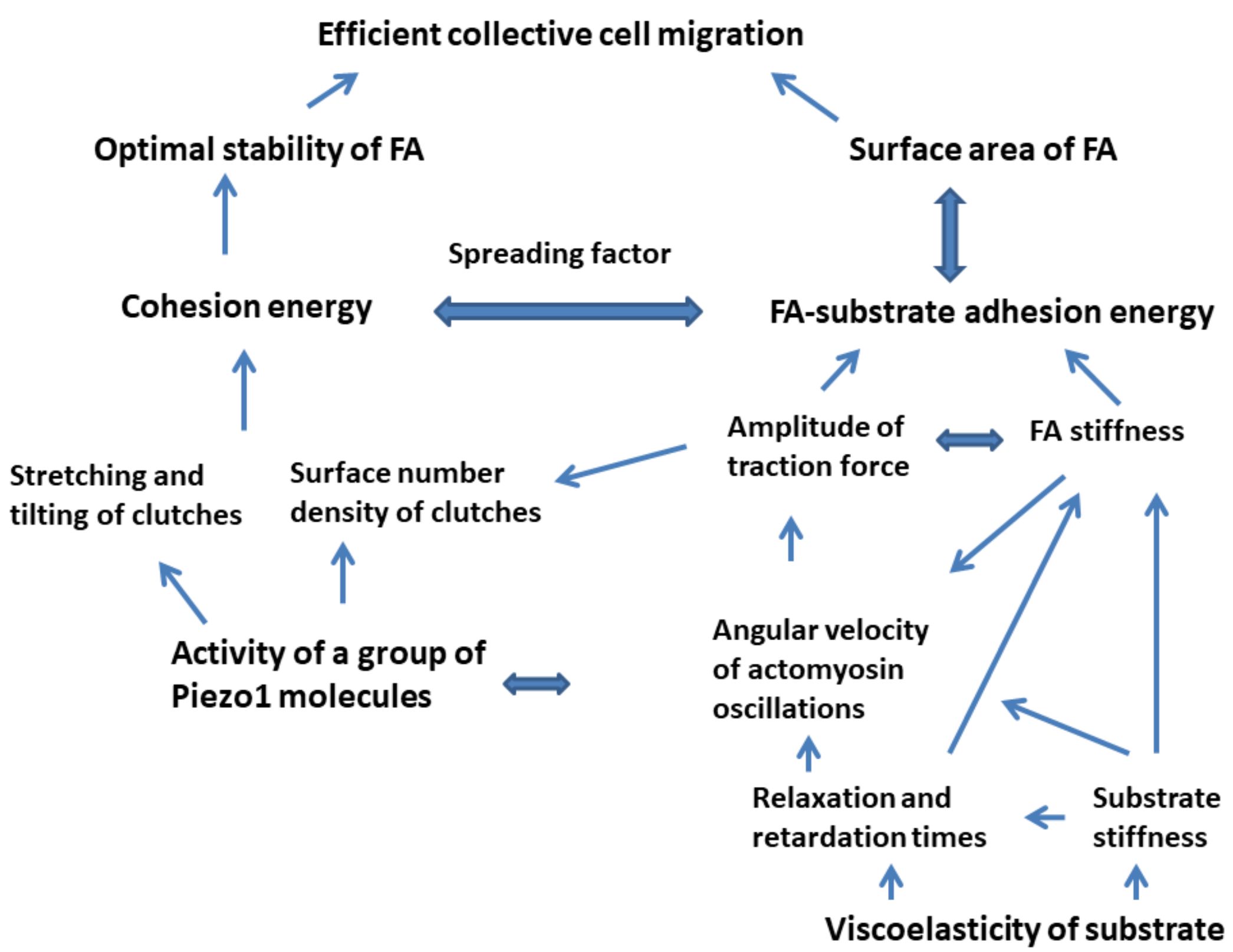

**Figure 3**. The relationships between physical parameters governing an FA's coupling with a viscoelastic substrate. The remodelling of FA can be regulated by altering the stiffness of the substrate and relaxation and retardation times.

The FA tend to adapt its state to the substrate's viscoelasticity by modulating its spring constant $k_{FA}$ towards the optimal value, which is biased toward an optimal range i.e., $k_{FA}(r,\tau) \to k_{FA}^{*}(r)$ ensuring maximum force amplitude $F_0^{max}(r)$. FAs exhibit enhanced stability on stiffer substrates and reach a larger steady-state surface area $A_{ss}$. For efficient spreading of FAs, it is necessary that: (i) the maximum force amplitude $F_0^{max} > F_c$ and (ii) that $\omega_{max}$ be low enough to ensure longer periods of traction force $F(t,\tau)$ and optimal activity of Piezo1 channels. The surface area of an FA at steady state depends on the parameters of substrate matrix such as the relaxation and retardation times $\tau_m$ and $\tau_u$, as well as on the spring constant $k_m$. The ratio between relaxation and retardation times $\frac{\tau_m}{\tau_u} = K_\tau$ (where $K_\tau$ is a constant that depends on the stiffness of substrate and satisfies the conditions that $K_\tau < 1$). Consequently, the relaxation and retardation time increase or decrease simultaneously. The main differences in FA-substrate coupling in soft and stiff viscoelastic substrates are shown in **Table1**:

**Table 1**. The main differences in FA-substrate coupling in soft and stiff viscoelastic substrates

| Physical parameters | Soft-to-stiff substrates |
|---|---|
| Relaxation and retardation times of substrate | $(\tau_u - \tau_m)^{soft} < (\tau_u - \tau_m)^{stiff}$ |
| Spring constants of FA and substrate | $(k_{FA}^* - k_m)^{soft} < (k_{FA}^* - k_m)^{stiff}$ |
| Force amplitude | $F_0^{soft} < F_0^{stiff}$ |
| Surface tension of FA | $\gamma_{FA}^{soft} < \gamma_{FA}^{stiff}$ |
| Quasi-steady surface area of FA | $A_{ss}^{soft} < A_{ss}^{stiff}$ |

An increase in the intracellular concentration of calcium in the optimal range, caused by the activity of Piezo1 channels, increases: (i) the amplitude of actomyosin oscillations, (ii) the amplitude of the traction force, (iii) the surface density of clutches, and (iv) the cohesion of FAs.

An increase in the relaxation and retardation time on a soft substrate from several tens of seconds to several hundreds of seconds stabilizes an FA and increases its cohesion as quantified by a higher surface tension $\gamma_{FA}$ [4]. Sharma et al. [4] highlighted that the FAs of HT-1080 fibrosarcoma reach optimal stability on soft, rapidly relaxing alginate–Reconstituted Basement Membrane (rBM) interpenetrating network (IPN) substrates, characterized by a Young's modulus of $E_m = 2\ \mathrm{kPa}$ and a relaxation time of $\tau_m = 10\ \mathrm{s}$, which facilitates their effective migration, a phenomenon referred to as super-diffusion. However, a long-relaxation time of $\tau_m = 1000\ \mathrm{s}$ causes the formation of more stable FAs leading to the damped cell migration described as sub-diffusion. Consequently, a shorter relaxation time $\tau_m$ promotes cell migration [4,62]. The maximum angular velocity $\omega_{max}$ is higher when the relaxation and retardation time are shorter, ensuring Piezo1-induced optimal stability of FAs even on very soft substrates [2].

A decrease in the relaxation and retardation time leads to: an increase in the maximum angular velocity $\omega_{max}$ toward the optimal value for Piezo1 activation and an increase in the force amplitude $F_0$ (high angular velocity regime) leading to an increase in the FA's quasi-steady state surface area $A_{ss}$. When the relaxation time $\tau_m$ increases from $\tau_m = 15 - 20\ \mathrm{s}$ to $\tau_m = 150 - 200\ \mathrm{s}$ on soft PEG-based substrates (Young's modulus of $E_m = 0.4\ \mathrm{kPa}$), the FA surface area of mesenchymal stem cells decreases from $A_{ss} \sim 1.1 - 1.3\ \mu\mathrm{m}^2$ to $A_{ss} \sim 0.6 - 0.7\ \mu\mathrm{m}^2$ [2]. In contrast to softer substrates, their steady surface area is less sensitive to changes in the relaxation and retardation times. It is in accordance with the fact that: (i) clutches establish more stable bonds to substrates and (ii) the tilting of clutches is reduced due to the reduced mobility of polymer chains. The FA ensures a larger quasi-steady surface area equal to $A_{ss} \sim 1.4 - 1.6\ \mu\mathrm{m}^2$ on stiff PEG-based substrates (Young's modulus of $E_m = 25\ \mathrm{kPa}$) for short relaxation times of $\tau_m = 20 - 30\ s$ and long relaxation times $\tau_m = 200 - 300\ \mathrm{s}$ [2]. The larger surface areas $A_{ss}$ on stiffer substrates are caused by larger force amplitudes $F_0$. Ciccone et al. [3] considered the steady-state FA surface areas of epithelial MCF-10A cells on slow- and fast- relaxing, soft and stiff micro-patterned PAAm substrates with Young's moduli of $0.4\ \mathrm{kPa}$ and $25\ \mathrm{kPa}$. FAs reach higher quasi-steady surface areas $A_{ss}$ on fast-relaxing than on slow-relaxing soft substrates. The surface area $A_{ss}$ is higher on stiff substrates and less sensitive to change in the relaxation time $\tau_m$ [3]. The magnitude of $A_{ss} = 1 - 2.5\ \mu m^2$ achieved on a fast-relaxing soft substrate promotes efficient cell migration, while the migration of cells is reduced on stiffer substrates due to the larger and more stable FAs [3].

Quantitative validation of the present framework requires integrated measurements of focal adhesion dynamics, traction forces, substrate viscoelasticity, cytoskeletal activity, and Piezo1-mediated signalling obtained from the same cell–matrix system under controlled conditions. Although individual parameters have been measured experimentally, a unified dataset enabling direct parameter identification and model testing is currently lacking.

## 5. Conclusion

The stability and size of focal adhesions (FAs), and consequently the efficiency of collective cell migration, are strongly influenced by the physical properties of viscoelastic substrates, including their stiffness and characteristic relaxation and retardation times. By modulating these parameters, substrate mechanics constrain FA behavior through their effects on effective adhesion stiffness, actomyosin dynamics, molecular clutch engagement, and mechanosensitive pathways such as Piezo1 activation. Within this context, FA dynamics can be naturally interpreted through the lens of active wetting and de-wetting processes.

A growing body of experimental and theoretical work supports a two-timescale description of cell–matrix interactions, in which fast actomyosin-driven force generation is coupled to slower adhesion remodelling. At short timescales, viscoelastic models such as the standard linear solid (Zener) framework predict that the transfer of mechanical work from the cytoskeleton to the substrate depends sensitively on the interplay between elastic energy storage and viscous dissipation. This interplay can give rise to optimal dynamical regimes—often described as resonance-like—where force transmission and traction generation are maximized. Such regimes are expected to promote coordinated activation of mechanosensitive pathways, including clustered Piezo1 activity and localized calcium signalling.

At longer timescales, FAs exhibit adaptive behaviour through force-dependent protein recruitment and structural remodelling. High-resolution measurements, including traction force microscopy and FRET-based tension sensors for proteins such as talin and vinculin, indicate that adhesions not only grow in size but also reorganize internally to match mechanical loading conditions. The recruitment of scaffold proteins such as paxillin and kindlin further stabilizes adhesion complexes under force, suggesting that FA maturation reflects an active tuning process. Within this framework, adhesion size and stability can be interpreted as the outcome of a balance between integrin–ligand adhesion energy and internal cohesion, consistent with a wetting/de-wetting analogy.

Importantly, substrate relaxation and retardation times emerge as key regulators of adhesion dynamics. Fast-relaxing substrates tend to support dynamic yet stable adhesions that enable efficient migration even in mechanically compliant environments, whereas slow-relaxing substrates can promote excessive stabilization and reduced motility. These trends are consistent with experimentally observed transitions between distinct migration regimes, including super-diffusive and sub-diffusive behaviour.

Taken together, these findings point toward timescale matching between cellular activity and material response as a central organizing principle in cell–matrix interactions. By integrating viscoelastic substrate mechanics, active force generation, and adhesion energetics, this perspective provides a unifying framework for understanding how cells regulate force transmission, adhesion stability, and mechanosensitive signalling.

Looking forward, several open questions remain. In particular, how cells dynamically tune their internal timescales in response to heterogeneous and evolving microenvironments, how stochastic fluctuations and spatial heterogeneities influence resonance-like behaviour, and how mechanical and biochemical feedbacks are integrated across scales are all areas that require further investigation. In this review article we proposed a new theoretical model describing these phenomena using the active wetting and de-wetting of FAs on viscoelastic substrates as a paradigm for developing predictive models of cell behaviour in complex, physiologically relevant settings.

**Conflict of interest**: The authors report there is no conflict of interest.

**Funding**: This work was supported in part by the Engineering and Physical Sciences Research Council, United Kingdom (grant number EP/X004597/1 and grant number EP/X033554/1), by the Ministry of Science, Technological Development and Innovation of the Republic of Serbia (Contract No. 451-03-34/2026-03/ 200135), the National Health and Medical Research Council of Australia (Investigator Grant L3 2034293).

## References

1. Hynes RO. (2002). Integrins: Bidirectional, allosteric signaling machines. Cell, 2002; 110(6):673-687, https://doi.org/10.1016/S0092-8674(02)00971-6.
2. Oliva-Armengol MAG, Ciccone G, Fläschner G, Luo J, Voigt JL, Romani P, Genever P, Dobre O, Dupont S, Vassalli M, Roca-Cusachs P, Salmeron-Sanchez M. Piezo1 regulates the mechanotransduction of soft matrix viscoelasticity. Nature Comm. 2025; 16(1):9155, doi: 10.1038/s41467-025-64185-5.
3. Ciccone G, Azevedo Gonzalez-Oliva M, Versaevel M, Cantini M, Vassalli M, Salmeron-Sanchez M, Gabriele S. Epithelial cell mechanoresponse to matrix viscoelasticity and confinement within micropatterned viscoelastic hydrogels. Adv. Sci. 2025; 12(18):2408635, https://doi.org/10.1002/advs.202408635.
4. Sharma V, Adebowale K, Gong Z, Chaudhuri O, Shenoy VB. Glassy adhesion dynamics govern transitions between sub-diffusive and super-diffusive cancer cell migration on viscoelastic substrates. Nature Comm. 2026: 17:978. https://doi.org/10.1038/s41467-025-67709-1.
5. Chen CS, Mrksich J, Huang S, G. Whitesides GM, Ingber DE. Geometric control of cell life and death, Science 1997; 276:1425–1428, DOI: 10.1126/science.276.5317.1425.
6. Engler AJ, Sen S, Sweeney HL, Discher DE. Matrix elasticity directs stem cell lineage specification, Cell 2006; 126:677–689, DOI: 10.1016/j.cell.2006.06.044.
7. Ziebert F. and Aranson IS. Spatiotemporal feedback between actomyosin and focal-adhesion systems optimizes rapid cell migration, PLoS ONE 2013; 8(3):e59634, https://doi.org/10.1371/journal.pone.0064511.
8. Chaudhuri O, Gu L, Klumpers D, Darnell M, Bencherif SA, Weaver JC, Huebsch N, Lesher-Perez SC, Mooney DJ. Hydrogels with tunable stress relaxation for 3D cell culture and stem cell biology. Nature Mat. 2016; 15(3):326–334, DOI: 10.1038/nmat4489.
9. Li Y and Chaudhuri O. Mechanical regulation of cell function by viscoelasticity, Trends Cell Biol. 2020; 30:1–14, doi: 10.1038/s41586-020-2612-2.
10. Chan CE and Odde DJ. Traction dynamics of filopodia on compliant substrates. Science 2008; 322(5908):1687-1691, DOI: 10.1126/science.1163595.
11. Yao M, Tijore A, Cheng D, Li JV, Hariharan A, Martinac B, Tran Van Nhieu G, Cox CD, Sheetz M. Force- and cell state–dependent recruitment of Piezo1 drives focal adhesion dynamics and calcium entry. Science Adv. 2022; 8(45):1461, doi:10.1126/sciadv.abo1461.
12. Pajic-Lijakovic I, Milivojevic M, Martinac B., McClintock PVE. Marangoni-Driven Redistribution and Activity of Piezo1 Molecules in Epithelial and Cancer Cells. Adv. Coll. Int. Sci. 2026; 353:103877, DOI: 10.1016/j.cis.2026.103877.
13. Pathak MM, Nourse JL, Tran T, Hwe J, Arulmoli J, Le DT, Bernardis E, Flanagan LA, Medha M. Stretch-activated ion channel Piezo1 directs lineage choice in human neural stem cells. PNAS 2014; 111(45):16148–16153, DOI: 10.1073/pnas.1409802111.
14. Bruinsma R and Sackmann E. (2001). Bioadhesion and the dewetting transition. Comptes Rendus de l'Académie des Sciences - Series IV - Physics-Astrophysics 2001 ; 2(6) :803–815, https://doi.org/10.1016/S1296-2147(01)01225-2.

15. de Gennes PG, Wetting: statics and dynamics, Rev. Mod. Phys. 1985; 57:827–863, https://doi.org/10.1103/RevModPhys.57.827.
16. Pérez-González C, Alert R, Blanch-Mercader C, Gómez-González M, Kolodziej T, Bazellieres E, Casademunt J, Trepat X. Active wetting of epithelial tissues. Nature Phys. 2019; 15:79-88, DOI: 10.1038/s41567-018-0279-5.
17. Pajic-Lijakovic, I and Milivojevic M. Active wetting of epithelial tissues: modeling considerations. Eur. Biophys. J. 2023; 52:1-15, DOI: 10.1007/s00249-022-01625-w.
18. Pajic-Lijakovic I, Milivojevic M, McClintock PVE. Anisotropy and shear stress accumulation during collective migration of epithelial cells. Europ. Biophys. J. 2026; 55:21-39, doi: 10.1007/s00249-026-01813-y.
19. Sheetz PM. Cell control by membrane–cytoskeleton adhesion, Nat. Rev. Mol. Cell Biol. 2001; 2:392–396, DOI: 10.1038/35073095.
20. Swaminathan V and Waterman CM. The molecular clutch model for mechanotransduction evolves. Nature Cell Biol. 2016; 18(5):459–461, https://doi.org/10.1038/ncb3350.
21. Lemke SB, Weidemann T, Cost AL, Grashoff C, Schnorrer F. A small proportion of Talin molecules transmit forces at developing muscle attachments in vivo. PLOS Biol. 2019; 17(3):e3000057, doi:10.1371/journal.pbio.3000057.
22. Stutchbury B, Atherton P, Tsang R, Wang DY, Ballestrem C. Distinct focal adhesion protein modules control different aspects of mechanotransduction. J. Cell Sci. 2017; 130(9), 1612–1624, DOI: 10.1242/jcs.195362.
23. Case LB, Baird MA, Shtengel G, Campbell SL, Hess HF, Davidson MW, Waterman CM. Molecular mechanisms of vinculin activation and organization at focal adhesions. Nature Cell Biol. 2015; 17(7), 880–892, DOI: 10.1038/ncb3180.
24. Yao M, Goult BT, Chen H, Cong P, Sheetz MP, Yan J. Mechanical activation of vinculin binding to talin locks talin in an unfolded conformation. Sci. Rep. 2014; 4:4610, doi:10.1038/srep04610.
25. Carisey A, Tsang R, Greiner AM, Nijenhuis N, Heath N, Nazgiewicz A, Buckley R, Enchev R, Goldmann WH, Guest B, Kechagia JZ, Anderson KI, Humphries JD, Humphries MJ.Vinculin regulates the recruitment and release of core focal adhesion proteins in a force-dependent manner. Current Biol. 2013; 23(4):271–281, DOI: 10.1016/j.cub.2013.01.009.
26. Bachir AI, Horwitz AR, Nelson WJ, Schleicher SB. Integrin-associated complexes form hierarchically with variable stoichiometry during nascent adhesion formation. Current Biol. 2014; 24(16):1845–1853, DOI: 10.1016/j.cub.2014.07.011.
27. Liu X, Zhang M, Wang P, Zheng K, Wang X, Xie W, Pan X, Shen R, Liu R, Ding J, Wei Q. Nanoscale distribution of bioactive ligands on biomaterials regulates cell mechanosensing through translocation of actin into the nucleus. PNAS 2025; 122(11):e2501264122, https://doi.org/10.1073/pnas.2501264122.
28. Moser M, Legate KR, Zent R, Fässler R. The tail of integrins, talin, and kindlins. Science 2009; 324(5929):895–899, DOI: 10.1126/science.1163865.
29. Lele TP, Pendse J, Kumar S, Salanga M, Karavitis J, Ingber DE. Mechanical forces alter zyxin unbinding kinetics in focal adhesions. Biophys. J. 2008; 94(6):2325–2333, DOI: 10.1002/jcp.20550.

30. Theodosiou M, Widmaier M, Brunner CA, Rakic V, Bauer MS, Müller DJ, Fässler R. Kindlin-2 cooperates with talin to activate integrins and recruit paxillin to focal adhesions. eLife 2016; 5,:e10134, DOI: 10.7554/eLife.10130.
31. Gingras AR, Bate N, Goult BT, Hazelwood L, Canestrelli I, Grossmann JG, Patel B, PryardJR, Wilkins R, Barsukov IL, Critchley DR. Central region of talin has a unique fold that binds vinculin and actin. J. Biol. Chem. 2010; 283(44):29905–29912, DOI: 10.1074/jbc.M109.095455.
32. Zhang X, Jiang G, Cai Y, Monkley SJ, Critchley DR, Sheetz MP. Talin dimerization is required for integrin activation and focal adhesion formation. J. Biol. Chem. 2008; 291(11):5589–5600, DOI: 10.1038/ncb1765.
33. del Rio A, Perez-Jimenez R, Liu R, Roca-Cusachs P, Fernandez JM, Sheetz MP. Stretching single talin rod molecules activates vinculin binding. Science 2009; 323(5914):638–641, DOI: 10.1126/science.1162912.
34. Wang X and Ha T. Defining single molecular forces required to activate integrin and notch signaling. Science 2013; 340(6135):991–994, DOI: 10.1126/science.1231041.
35. Pajic-Lijakovic I, Milivojevic M., Martinac B, McClintock PVE. Targeted elimination of mesenchymal-like cancer cells through cyclic stretch activation of Piezo1 channels: the physical aspects. Biophys. Rev. 2025; 17: 847–865, DOI: 10.1007/s12551-025-01304-y.
36. Pajic-Lijakovic I, Milivojevic M., Martinac B, McClintock PVE. Irregular Curvature at Focal Adhesions Modulates Piezo1 Activity and Low-Frequency Ultrasound–Induced Apoptosis in Cancer Cells. Phys Life Rev 2026; 58:36-53, DOI: 10.1016/j.plrev.2026.06.004.
37. Verkest C, Roettger L, Zeitzschel N, Hall J, Sánchez-Carranza O, Huang AT-L, Lewin GR, Lechner S. G. Cluster nanoarchitecture and structural diversity of PIEZO1 at rest and during activation in intact cells. Nature Comm. 2024; 15(1):1251, DOI: 10.1126/sciadv.ady8052.
38. Syeda R, Florendo MN, Cox CD, Kefauver JM, Santos JS, Martinac B, Patapoutian A. Piezo1 Channels Are Inherently Mechanosensitive, Cell Rep. 2016; 17(7):1739-1746, https://doi.org/10.1016/j.celrep.2016.10.033.
39. Franco SJ, Rodgers MA, Perrin BJ, Tsuruta H, Bennin DA, Hoelzle MK, Huttenlocher A. Calpain-mediated proteolysis of talin regulates adhesion dynamics. Nature Cell Biol. 2004; 6(10):977–983, DOI: 10.1038/ncb1175.
40. Javed A, Stubb A, Villeneuve C, Myllymäki SM, Peters F, Rübsam M, Niessen CM, Biggs LC, Wickström SA. Piezo1 balances membrane and cortex tension to stabilize intercellular junctions and maintain the epithelial barrier. J. Cell Sci. 2025; 138(16):jcs263938, DOI: 10.1242/jcs.263938.
41. Coyer SR, Singh A, Dumbauld DW, Calderwood DA, Craig SW, Delamarche E, García AJ. Nanopatterning reveals an ECM area threshold for focal adhesion assembly and force transmission that is regulated by integrin activation and cytoskeleton tension. J. Cell Sci. 2012; 125(21):5110–5123, https://doi.org/10.1242/jcs.108035.
42. Sen S, Engler AJ, Discher DE. Matrix strains induced by cells: mechanics and mechanobiology, Biophys. J. 2008; 95:2193–2206, DOI: 10.1007/s12195-009-0052-z.
43. Gardel ML, Sabass M, Ji L, Danuser G, Schwarz US, Waterman CM. Traction stress in focal adhesions correlates biphasically with actin retrograde flow. Cell 2008; 135:1255–1267, DOI: 10.1083/jcb.200810060.

44. Banerjee DS, Munjal A, Lecuit T, Rao M. Actomyosin pulsation and flows in an active elastomer with turnover and network remodeling. Nature Comm. 2017; 8(1):1121, doi:10.1038/s41467-017-01130-1.
45. Hecht I, Bar-Ziv R, Safran, SA. Actin-myosin-based vascular boundary formation. PNAS 2015; 112(46):14150-14155. doi:10.1073/pnas.1511211112.
46. Sumigray KD and Lechler T. Cell Adhesion and the Actin Cytoskeleton in Epithelial Morphogenesis. Curr. Topics Dev. Biol. 2015; 112:383-414, doi:10.1016/bs.ctdb.2014.11.012.
47. Margadant F, Chew LL, Hu X, Kohiat H, Sheetz MP, Grenci G. Mechanotransduction in vivo by repeated talin stretch-relaxation events depends upon vinculin. PLOS Biol. 2011; 9(12):e1001223, https://doi.org/10.1371/journal.pbio.1001223.
48. Han SJ, Goult BT, Stehbens SJ, Waterman CM. Talin–vinculin pre-complexation serves as a node for enhancing nascent adhesion maturation. eLife 2021; 10:e66151, doi: 10.7554/eLife.66151.
49. Golji J, Lam J, Mofrad MRK. Vinculin activation is necessary for complete talin binding. Biophys. J. 2011; 100(2):332–340, DOI: 10.1016/j.bpj.2010.11.024.
50. Chakraborty S, Chaudhuri D, Banerjee S, Bhatt M, Haldar S. Direct observation of chaperone-modulated talin mechanics with single-molecule resolution. Comm. Biol. 2022; 5(1):307, https://doi.org/10.1038/s42003-022-03258-3.
51. Miao H, Blankenship JT. The pulse of morphogenesis: actomyosin dynamics and regulation in epithelia. Development 2020; 147(17):dev186502. doi:10.1242/dev.186502.
52. Ghosh D, Ghosh S, Chaudhuri A. Deconstructing the role of myosin contractility in force fluctuations within focal adhesions. Biophys. J. 2022; 121(10):1851-1863, doi: 10.1016/j.bpj.2022.03.025.
53. Yao M, Goult BT, Klapholz B, Hu X, Toseland CP, Guo Y, Cong P, Sheetz MP, Yan J. The mechanical response of talin. Nature Comm. 2016; 7(1):11966, DOI: 10.1038/ncomms11966.
54. Lin CY. Physical interpretation and essence of the standard linear solid model. International J. Mech. Sci. 2025; 310:111139, https://doi.org/10.1016/j.ijmecsci.2025.111139.
55. Bronshtein T, Au-Yeung GCT, Sarig U, Nguyen EBV, Mhaisalkar PS, Boey FYC, Venkatraman SS, Machluf M. A Mathematical Model for Analyzing the Elasticity, Viscosity, and Failure of Soft Tissue: Comparison of Native and Decellularized Porcine Cardiac Extracellular Matrix for Tissue Engineering. Tissue Eng. Part C: Meth. 2013; 19(8):620–630, doi: 10.1089/ten.tec.2012.0387.
56. Schwarz US, Balaban NQ, Riveline D, Bershadsky A, Geiger B, Safran SA. Calculation of forces at focal adhesions from elastic substrate data: the effect of localized force and the need for regularization. Biophys. J. 2002; 83(3):1380–1394, https://doi.org/10.1016/S0006-3495(02)73909-X.
57. Lewis AH, Cui AF, McDonald MF, Grandl, J. Transduction of Repetitive Mechanical Stimuli by Piezo1 and Piezo2 Ion Channels. Cell Rep. 2017; 19(12):2572–2585, http://creativecommons.org/licenses/by-nc-nd/4.0/.
58. Kanchanawong P, Shtengel G, Pasapera AM, Ramko EB, Davidson MW, Hess HF, Waterman CM. Nanoscale architecture of integrin-based cell adhesions. Nature 2010; 468(7323):580–584, DOI: 10.1038/nature09621.

59. Burgos-Bravo F, Figueroa NL, Casanova-Morales N, Quest AFG, Wilson CAM, Leyton L. (2018). Single-molecule measurements of the effect of force on Thy-1/integrin interaction using nonpurified proteins. Mol. Biol. Cell 2018; 29(10):1211–1221, DOI:10.1091/mbc.E17-03-0133.
60. Myers KR and Freed KF. Surface tension of dilute polymer solutions. II. The second virial coefficient. J Chem Phys. 1993; 98:2437-2450, https://doi.org/10.1063/1.464171.
61. Pajic-Lijakovic I, Eftimie R, Milivojevic M, Bordas SPA. Multi-scale nature of the tissue surface tension: theoretical consideration on tissue model systems. Adv. Coll. Int. Sci. 2023; 315:102902, DOI: 10.1016/j.cis.2023.102902.
62. Chaudhuri O, Gu L, Darnell M, Klumpers D, Bencherif SA, Weaver JC, Huebsch N, Mooney DJ. (2015). Substrate stress relaxation regulates cell spreading. Nature Comm. 2015; 6:6365, https://doi.org/10.1038/ncomms7365.